\documentclass[]{article}
\usepackage{amsmath}
\usepackage{amsfonts}
\usepackage{amssymb}
\usepackage[symbol]{footmisc}
\usepackage{hyperref}
\usepackage{psfrag}
\usepackage{graphicx}
\newcommand{\runninghead}[2]{}
\newcommand{\address}[1]{}

\newcommand{\pd}[2]{\ensuremath{\frac{\partial #1}{\partial #2}}}
\newcommand{\pdl}[2]{\ensuremath{\partial #1 / \partial #2}}

\newcommand{\ie}{\textit{i.e.},~}
\newcommand{\eg}{\textit{e.g.},~}

\newcommand{\vs}{\ensuremath{v_\text{s}}}

\newcommand{\rs}{\ensuremath{\rho_\text{s}}}

\newcommand{\rn}{\ensuremath{\rho_\text{n}}}
\newcommand{\vn}{\ensuremath{v_\text{n}}}

\newcommand{\vsb}{\ensuremath{\mathbf{v}_\text{s}}}
\newcommand{\vnb}{\ensuremath{\mathbf{v}_\text{n}}}

\newcommand{\E}{\mathrm{e}}
\newcommand{\I}{\mathrm{i}}

\DeclareMathOperator{\Imp}{Im}
\author{L.A.\,Melnikovsky\footnote{E-mail: leva@kapitza.ras.ru}}
\title{On Sound Reflection in Superfluid}
\date{}
\begin{document}
\psfragscanon
\maketitle
\begin{abstract}
%
We consider reflection of the first and the second sound waves by a
rigid flat wall in superfluid. Nontrivial dependence of the reflection
coefficients on the angle of incidence is obtained. Sound conversion is
predicted at slanted incidence.
\end{abstract}

\section{Introduction}
Bulk superfluid has twice as many variables as a normal
fluid\cite{landau}. Direct consequence is the existence of two
independent sound modes in helium~II referred to as the first sound and
the second sound. Due to anomalously small thermal expansion of helium,
these modes can be viewed as purely pressure and temperature waves
respectively. The ``purity'' is essential, \eg for the effectiveness of
sound emission by various sources.

Restricted geometry effectively eliminates some of the hydrodynamic
variables. Particularly, a steady cell wall eliminates the
normal component of the mass flux $\mathbf{j}$ and the
normal velocity
\vnb: the boundary condition at the wall is
\begin{equation}
\label{bound}
\mathbf{j}_\perp=0, \quad \vnb=0. 
\end{equation}
The fourth sound \cite{atkins} is one result of such elimination. This
is the only sound mode in a narrow channel, both temperature and
pressure oscillate coherently in this wave. In general, the sound modes,
independent in bulk liquid, begin to interact at the
boundary.\footnote{Sound reflection at the free helium surface
and at the solid helium boundary is extensively explored
\cite{pellamt},\cite{dingle}, \cite{chernikova}, and \cite{kagkos}.}
This necessitates complete two-fluid consideration of
the sound reflection in superfluid which is the subject of the present
paper.

To avoid complications associated with the wall deformation consider a
perfectly rigid flat wall. We therefore ignore numerous peculiarities of
the sound transmission into solids and limit ourselves to linear
hydrodynamic equations only. In Sec.\ref{harm} we find all {\em three}
nontrivial harmonic (\ie proportional to $\E^{\I\mathbf{kr} - \I\omega
t}$) solutions of the equations. It is important to mention that
unbounded solutions (those with complex wave vector $\mathbf{k}$) should
not be overlooked in the restricted geometry.

The boundary condition
\eqref{bound} selects a two-dimensional subspace of solutions
for particular frequency $\omega$. Specific solutions correspond to
the first sound reflection (Sec.\ref{first}) and the second
sound reflection (Sec.\ref{second}). Interestingly enough,
some part of the incident wave energy is transformed between
the first and the second sound.


\section{Harmonic solutions}
\label{harm}
Consider the linearized equations of superfluid hydrodynamics\cite{khalat}:
\begin{gather}
\label{h1}
\dot{\rho} + \pd{j^i}{x^i} =0,\\
\label{h2}
\dot{j^i} + \pd{p}{x^i} =
\eta \pd{}{x^k}\left(\pd{\vn^i}{x^k}+\pd{\vn^k}{x^i}-\delta^{ik}\frac{2}{3}\pd{\vn^l}{x^l}\right)
+\pd{}{x^i}\left(\zeta_2\pd{\vn^l}{x^l}-\zeta_1 \pd{\rs w^l}{x^l}
\right),\\
\label{h3}
\dot{\vs^k} + \pd{\mu}{x^k} =
\pd{}{x^k} \left(
\zeta_4 \pd{\vn^l}{x^l}
-\zeta_3 \pd{\rs w^l}{x^l} 
\right),\\
\label{h4}
T\left(\dot{\sigma}\rho + \sigma\dot{\rho} +\sigma\rho\,\pd{\vn^l}{x^l}\right) =
\kappa\,\pd{^2T}{x^l \partial x^l},
\end{gather}
where $\eta$, $\zeta_1=\zeta_4$, $\zeta_2$, $\zeta_3$, $\kappa$ are
dissipative coefficients, $\sigma$ is entropy per unit mass, $p$, $\mu$, $T$ are pressure, chemical
potential, and temperature, \vsb, \vnb, and $\mathbf{w}=\vnb-\vsb$ are superfluid, normal, and relative
velocities, $\rho$ and $\mathbf{j}$ are mass and momentum
densities. The velocities and the momentum density are coupled by the equation
\begin{equation}
\label{relw}
\mathbf{j}=\rho\vsb+\rn \mathbf{w} = \rho\vnb - \rs \mathbf{w}.
\end{equation}

Further simplification is facilitated by ignoring thermal expansion (we
therefore disregard the difference between specific heats $c=T\pdl{\sigma}{T}$ at constant pressure and
at constant volume). Namely put
\begin{align}
\label{compress}
p' &= s^2 \rho',\\
\label{thermi}
\rho \mu'&=-\sigma\rho T'+ p' = -\sigma\rho T'+ s^2 \rho', \text{ and} \\
\label{capacit}
T \sigma' &= c T',
\end{align}
where $s=(\pdl{p}{\rho})^{1/2}$ is the first sound velocity. The prime denotes small deviation
of the variables from their equilibrium values.

In a harmonic perturbation, space and time dependence of all deviations
has a form of $\exp(\I\mathbf{kr} - \I \omega t)$. To find all possible
harmonic excitations in bulk superfluid we substitute this exponential
term in Eqs.\ref{h1}-\ref{h4} and keep linear terms only.

The mass conservation \eqref{h1} gives
\begin{equation}
\label{mass}
\omega \rho' = k^i j^i.
\end{equation}

The momentum conservation law \eqref{h2} can be transformed as follows:
\begin{equation*}
-\I\omega j^i + \I p' k^i =
- \eta k^k \left(\vn^i k^k + k^i\vn^k - \frac{2}{3}\delta^{ik} k^l\vn^l\right)
- k^i k^k \left(-\rs \zeta_1 w^k + \zeta_2 \vn^k \right),
\end{equation*}
and, using \eqref{compress}, \eqref{relw}, and \eqref{mass}
\begin{equation}
\label{eigen1}
 (\omega + \I \eta k^2/\rho ) j^i
+\left(\I A - s^2 / \omega\right) k^i k^k j^k
+ \I \eta k^2 \rs w^i /\rho
+\I B k^i k^k w^k=0,
\end{equation}
where the constants $A= \left(\eta/3+\zeta_2\right)/\rho$
and $B=(A-\zeta_1)\rs$.

From the energy conservation law \eqref{h4} for harmonic deviation
we get
\begin{equation*}
  T \omega \sigma' \rho 
+ T \omega \sigma  \rho'
- T \sigma \rho \vn^l k^l
+\I k^2 \kappa T'=0.
\end{equation*}
Using \eqref{relw}, \eqref{capacit}, and \eqref{mass} this can be reduced to
\begin{equation}
\label{temp}
T \sigma \rs k^i w^i
=(c \omega \rho +\I \kappa k^2)T'.
\end{equation}

Finally, substituting exponential term and \eqref{relw} in \eqref{h3} we obtain
\begin{equation*}
-\I \omega (j^i-\rn w^i)/\rho +\I k^i \mu' =
-k^i \left(-\rs \zeta_3 w^k k^k + \zeta_4 \left(j^k+\rs w^k\right)k^k/\rho\right)
\end{equation*}
Combining this with \eqref{thermi}, \eqref{mass}, and \eqref{temp}
\begin{equation}
\label{eigen2}
 \omega j^i 
+\left(\I \zeta_4- s^2/\omega\right) k^i k^k j^k
-\omega \rn w^i 
+G k^i k^k w^k 
=0,
\end{equation}
where
\begin{equation*}
G\approx
\frac{T \rs \sigma^2 }{c \omega} -
\I\rs \left( \frac{T \kappa k^2 \sigma^2 }{c^2 \omega^2 \rho} 
-(\zeta_4 - \rho \zeta_3)\right)
\end{equation*}

Eqs. \eqref{eigen1} and \eqref{eigen2} can be written together as
\begin{equation*}
\hat{L}
\begin{pmatrix}
\mathbf{j}\\
\mathbf{w}
\end{pmatrix}
=0,
\end{equation*}
where $\hat{L}$ is a square $6\times 6$ matrix composed of the
coefficients from \eqref{eigen1} and \eqref{eigen2}. The linear system is
consistent if $\det \hat{L}=0$. Due to the system isotropy,
the determinant can not depend on individual components of $k^i$. Instead it
depends on $k^2=k^i k^i$ only. We therefore can put $k^y=k^z=0$ and
treat $\hat{L}$ as a $4\times 4$ matrix:
\begin{equation*}
\hat{L}=
\begin{pmatrix}
\omega +\left(- s^2 / \omega + \I A + \I \eta /\rho \right) k^2 & 0 &
\I ( \eta \rs /\rho + B) k^2 & 0 \\
0 & \omega + \I \eta k^2/\rho &
0 & \I \eta k^2 \rs /\rho\\
\omega +\left(-s^2/\omega+ \I \zeta_4\right)k^2 & 0 &
-\omega \rn + G k^2 & 0 \\
0 & \omega &
0 & -\omega \rn
\end{pmatrix}
\end{equation*}
After factorization $\det \hat{L}$ simplifies to
\begin{equation*}
 \begin{vmatrix}
 \omega +\left( \I A + \I \eta /\rho - s^2 / \omega \right) k^2 &
 \I ( \eta \rs /\rho + B) k^2 \\
 \omega +\left(\I \zeta_4 -s^2/\omega\right)k^2 &
 -\omega \rn + G k^2
 \end{vmatrix}
\cdot
 \begin{vmatrix}
 \omega + \I \eta k^2/\rho &
 \I \eta k^2 \rs /\rho\\
 \omega &
 -\omega \rn
 \end{vmatrix}.
\end{equation*}
All nontrivial solutions immediately follow:
\begin{align}
\label{sol1}
\omega^2
&\approx k_1^2 \left( s^2 - \I\omega A - \I\frac{\omega \eta}{\rho} \right)
\approx k_1^2 \left(s^2 
-\I\frac{\omega}{\rho}\left(\frac{4\eta}{3}+\zeta_2\right)\right)
\approx
k_1^2 s^2,\\
\label{sol2}
\omega^2
&\approx k_2^2 \omega_2 ( G - \I \frac{\eta \rs}{\rho} -\I B)/\rn
\approx 
k_2^2 \frac{T \rs \sigma^2 }{c \rn} 
\equiv k_2^2 s_2^2,\\
\label{sol3}
\omega&=-\I \eta k_3^2 /\rn,
\end{align}
where $s_2$ is the second sound velocity.

Roots $\mathbf{k}_1$ \eqref{sol1} and $\mathbf{k}_2$ \eqref{sol2} correspond to
``longitudinal''
solutions where $j^i \propto w^i
\propto k_{1,2}^i$, while the root $\mathbf{k}_3$ \eqref{sol3} corresponds to a
``transverse'' one $j^i k_3^i = w^i k_3^i = 0$.
The approximation in \eqref{sol1} and \eqref{sol2} is based on an assumption of
low bulk damping, \ie $|k_3| \gg |k_2| > |k_1|$. This implies complete splitting
between the first and second sound, namely $\mathbf{w}=0$ for
\eqref{sol1} and $\mathbf{j}=0$ for \eqref{sol2}. In the third
solution \eqref{sol3}, the superfluid velocity vanishes $\vsb=0$, \ie
the mass flux and the relative velocity are coupled by
the relation
$\mathbf{j}= \rn \mathbf{w}$.

\section{First sound reflection}
\label{first}
\begin{figure}
\begin{center}
\includegraphics[scale=1]{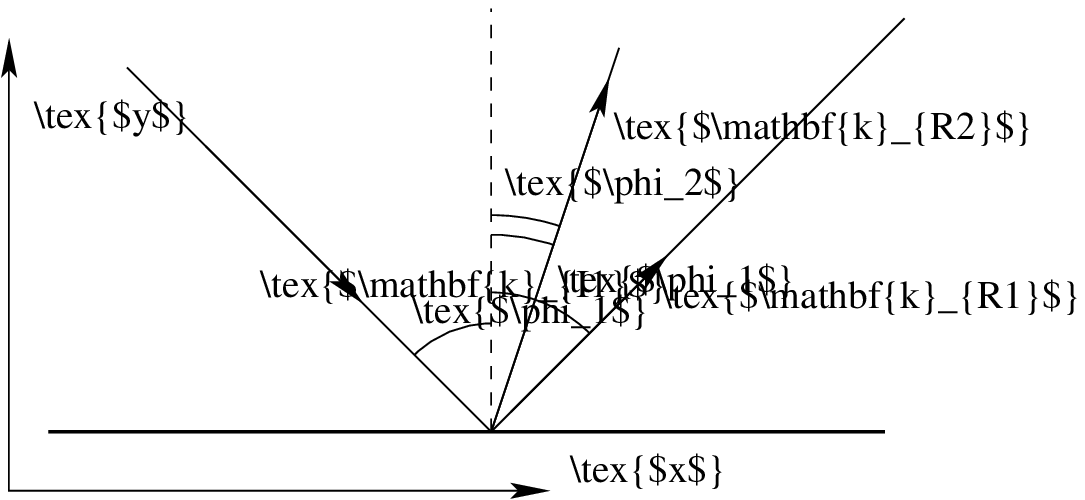}
\end{center}
\caption{First sound reflection}
\label{fig1}
\end{figure}

Consider the first sound wave incident towards the impervious rigid
plane wall at an angle $\phi_1$ (see Fig.\ref{fig1}). The subscripts
$I1$, $R1$, and $R2$ refer to the incident first, reflected first, and
reflected second sound waves respectively. The $x$ axis runs along the
wall and the $y$ axis is directed into the liquid.

The heat transfer through the
interface at low temperature can be neglected due to Kapitza resistance.
Appropriate boundary conditions are $j^y=0$, $w^y=0$, $j^x+\rs w^x =0$.
They can be written in the matrix form
\begin{equation}
\label{bound1}
\begin{pmatrix}
 j_{I1} \sin\phi_1 \\
 -j_{I1} \cos\phi_1 \\
 0 \\
 0
\end{pmatrix}
+
\begin{pmatrix}
 j_{R1} \sin\phi_1 \\
 j_{R1} \cos\phi_1 \\
 0 \\
 0
\end{pmatrix}
+
\begin{pmatrix}
 0 \\
 0 \\
 w_{R2} \sin\phi_2 \\
 w_{R2} \cos\phi_2
\end{pmatrix}
+
\begin{pmatrix}
\rn w_3^x\\
\rn w_3^y\\
w_3^x\\
w_3^y\\
\end{pmatrix}
=
\begin{pmatrix}
-\rs w\\
0\\
w\\
0
\end{pmatrix},
\end{equation}
where $\cos^2 \phi_2 = 1-(s_2^2/s^2)\sin^2\phi_1$,
to satisfy the condition $k_{I1}^x=k_{R2}^x$.
The last term in the left-hand side of \eqref{bound1}
represents the transverse surface wave with a wave vector
$\mathbf{k}_3$. 
The wave must decay away from the boundary, therefore
$\Imp k_3^y >0$. This requirement selects the sign in \eqref{w3},
which is the transversality relation $\mathbf{w}_3 \perp \mathbf{k}_3$:
\begin{equation*}
\label{w3}
\begin{pmatrix}
w_3^x\\
w_3^y 
\end{pmatrix}
 \propto
\begin{pmatrix}
 -k_3^y\\
 k_3^x 
\end{pmatrix}
 \approx
\begin{pmatrix}
 -k_3\\
 k_1\sin\phi_1
\end{pmatrix}
 =
\begin{pmatrix}
 -\varkappa \E^{\I \pi/4}\\
 k_1\sin\phi_1
\end{pmatrix}
=
\begin{pmatrix}
 -\varkappa \E^{\I \pi/4}\\
 k_2\sin\phi_2
\end{pmatrix},
\end{equation*}
where $\varkappa=\sqrt{\omega \rn/\eta}$.
Substituting this in \eqref{bound1} we get
\begin{equation*}
\begin{pmatrix}
 j_{I1} \sin\phi_1 \\
 -j_{I1} \cos\phi_1 \\
 0
\end{pmatrix}
+
\begin{pmatrix}
 j_{R1} \sin\phi_1 \\
 j_{R1} \cos\phi_1 \\
 0
\end{pmatrix}
+
\begin{pmatrix}
 \rs w_{R2} \sin\phi_2 \\
 0 \\
 w_{R2} \cos\phi_2
\end{pmatrix}
+
\begin{pmatrix}
\rho w_3^x\\
\rn w_3^y\\
w_3^y\\
\end{pmatrix}
=0
\end{equation*}
and
\begin{equation*}
2 j_{I1}
+w_{R2}
\left(
 \rs \frac{\sin\phi_2}{\sin\phi_1}
+\rn \frac{\cos\phi_2}{\cos\phi_1}
 + \E^{\I \pi/4} 
\frac{\varkappa \rho \cos\phi_2}{k_1 \sin^2\phi_1}
\right)
=0
\end{equation*}
Second sound is slower than the first one $s>s_2$, consequently $\phi_2<\pi/2$
and $\cos\phi_2 \neq 0$. One can therefore neglect the first term in parenthesis
\begin{equation}
\label{refl12}
w_{R2}=
- j_{I1}\frac{\sin^2\phi_1}{\cos\phi_2}
\frac{2 k_1 }{
 \rn k_1 \sin\phi_1 \tan\phi_1
 + \E^{\I \pi/4} 
\varkappa \rho
}.
\end{equation}
Similarly, the amplitude of the reflected first sound
is obtained from the equation
\begin{equation*}
\begin{pmatrix}
 j_{I1} \rn k_1\sin^2\phi_1 \\
 -j_{I1} \rho \varkappa \E^{\I \pi/4} \cos\phi_1
\end{pmatrix}
+
\begin{pmatrix}
 j_{R1} \rn k_1\sin^2\phi_1 \\
 j_{R1} \rho \varkappa \E^{\I \pi/4} \cos\phi_1
\end{pmatrix}
+
\begin{pmatrix}
\rho \rn k_1\sin\phi_1 w_3^x \\
\rho \rn\varkappa \E^{\I \pi/4} w_3^y
\end{pmatrix}
=0.
\end{equation*}
From this we have
\begin{equation}
\label{refl11}
\frac{j_{R1}}{j_{I1}}=\frac{\rho \varkappa
\cos\phi_1-\rn k_1\sin^2\phi_1 \E^{-\I\pi/4}}{\rho \varkappa
\cos\phi_1+\rn k_1\sin^2\phi_1 \E^{-\I\pi/4}}.
\end{equation}

\begin{figure}[ht]
\begin{center}
\psfrag{R11}[rb][rb]{$R_{11}$}
\psfrag{R12}[rt][rt]{$R_{12}$}
\psfrag{xlabel}[rt][rt]{$\phi_1$}
 \includegraphics[scale=0.7]{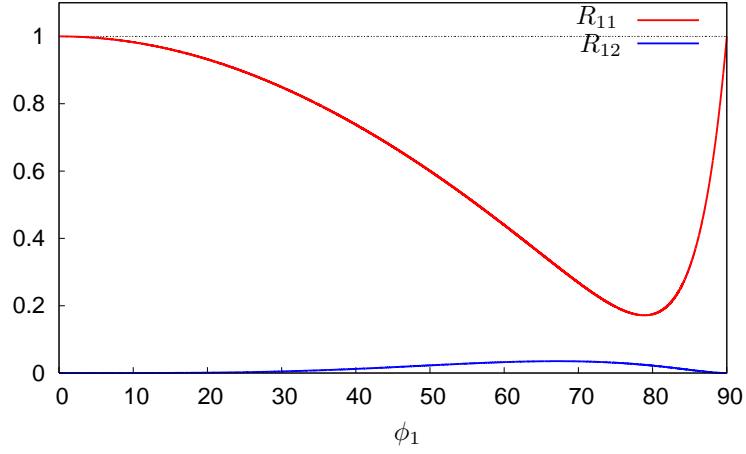}
\end{center}
\caption{Reflection and conversion coefficients
$R_{11}$ and $R_{12}$ \textit{vs.}
the angle of incidence $\phi_1$}.
\label{R1}
\end{figure}

Reflection and conversion efficiency must be characterized by
appropriate coefficients $R_{11}=F_{R1}/F_{I1}$ and
$R_{12}=F_{R2}/F_{I1}$ respectively. Here $F_1$ and $F_2$ are the energy fluxes
in the first and second sound waves. They are given by the expressions
\begin{equation*}
F_1=\frac{s}{2\rho} |j|^2,\quad
F_2=\frac{s_2\rs\rn}{2\rho} |w|^2.
\end{equation*}
Using \eqref{refl11} and \eqref{refl12} we get
\begin{align}
R_{11}&= \left|
	\frac{\rho \varkappa \cos\phi_1-\rn k_1\sin^2\phi_1 \E^{-\I\pi/4}}
             {\rho \varkappa \cos\phi_1+\rn k_1\sin^2\phi_1 \E^{-\I\pi/4}}
	\right|^2,\\
R_{12}&= 
\frac{s_2}{s}
\frac{\sin^4\phi_1}{\cos^2\phi_2}
\frac{ 4 \rs\rn k_1^2}
{\left|
\rn k_1 \sin\phi_1 \tan\phi_1
+ \rho\varkappa \E^{\I \pi/4} 
\right|^2}.
\end{align}
Sample graph of these functions is illustrated on Fig.\ref{R1}.
The reflection coefficient $R_{11}$ has a minimum of
\begin{equation}
\min R_{11}=3-2\sqrt{2}
\end{equation}
at finite angle of incidence.
The value at the minimum is the same as for the sound reflection
in usual hydrodynamics\cite{konst}.

\section{Second sound reflection}
\label{second}

\begin{figure}
\begin{center}
\includegraphics[scale=0.9]{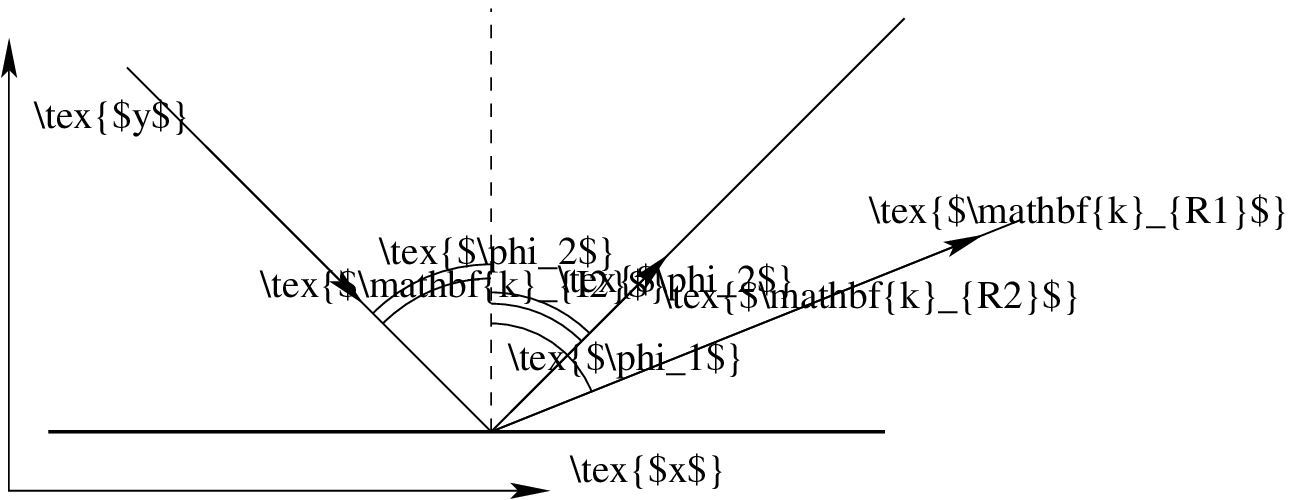}
\end{center}
\caption{Second sound reflection}
\label{fig2}
\end{figure}

The same approach can be used to investigate the second sound wave
incident at an angle $\phi_2$ (see Fig.\ref{fig2}). The boundary
conditions in this case are
\begin{equation*}
\begin{pmatrix}
 0 \\
 0 \\
 w_{I2} \sin\phi_2 \\
 -w_{I2} \cos\phi_2
 \end{pmatrix}
+
\begin{pmatrix}
 0 \\
 0 \\
 w_{R2} \sin\phi_2 \\
 w_{R2} \cos\phi_2
\end{pmatrix}
+
\begin{pmatrix}
 j_{R1} \sin\phi_1 \\
 j_{R1} \cos\phi_1 \\
 0 \\
 0
\end{pmatrix}
+
\begin{pmatrix}
\rn w_3^x\\
\rn w_3^y\\
w_3^x\\
w_3^y\\
\end{pmatrix}
=
\begin{pmatrix}
-\rs w\\
0\\
w\\
0
\end{pmatrix}.
\end{equation*}
After simplification this gives 
\begin{equation*}
2 \rs\rn w_{I2} 
=
j_{R1}
\left(
-\frac{\rho \varkappa \E^{\I\pi/4}\cos\phi_1}{k_2\sin^2\phi_2}
-\rn \frac{\sin\phi_1}{\sin\phi_2}
-\rs \frac{\cos\phi_1}{ \cos\phi_2}
\right)
\end{equation*}
The last term in parenthesis is always negligible (last equation is
meaningful only if $\sin\phi_2 < s_2 / s$). This gives
\begin{equation}
j_{R1}
=
-\rs w_{I2} 
\frac{2\rn k_2 \sin^2\phi_2}{\rho  \varkappa \E^{\I\pi/4}\cos\phi_1
+\rn k_2\sin\phi_1 \sin\phi_2}.
\end{equation}
The conversion coefficient $R_{21}=F_{R1}/F_{I2}$ is therefore given by
\begin{equation}
R_{21}=
\frac{4 s}{s_2}
\frac{\rs\rn k_2^2 \sin^4\phi_2}{\left|
\rho  \varkappa \E^{\I\pi/4}\cos\phi_1+\rn k_2\sin\phi_1 \sin\phi_2
\right|^2}.
\end{equation}
Its maximum
\begin{equation}
\max R_{21}=
4 \frac{\rs s_2}{\rn s}
\end{equation}
is reached at the critical angle $\sin\phi_2 = s_2 / s$.

\begin{figure}[ht]
\begin{center}
\psfrag{R22}[rb][rb]{$R_{22}$}
\psfrag{R21}[rt][rt]{$R_{21}$}
\psfrag{xlabel}[rt][rt]{$\phi_2$}
 \includegraphics[scale=0.7]{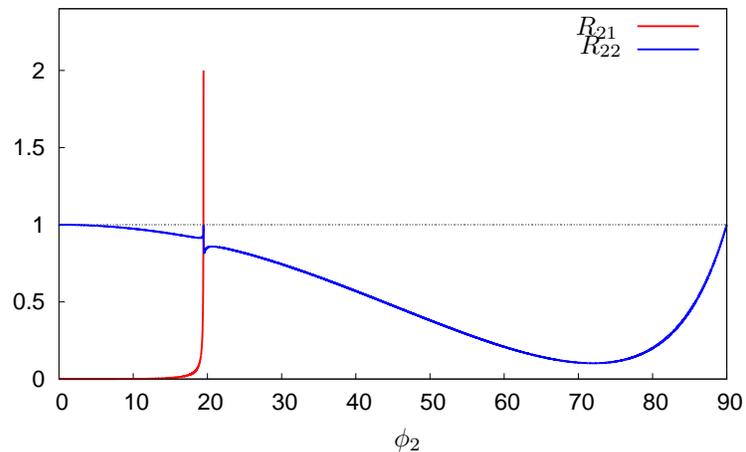}
\end{center}
\caption{Reflection and conversion coefficients
 $R_{22}$ and $R_{21}$ \textit{vs.} the angle of incidence $\phi_2$.}
\label{R2}
\end{figure}

Amplitude of the reflected second sound wave is found from the relation
\begin{equation}
\frac{w_{R2}}{w_{I2}}
\left( \frac{\varkappa \rho \E^{\I\pi/4}}{k_2 \sin\phi_2}
+\rn \tan\phi_1 +\rs \tan \phi_2\right)
=
\left( \frac{\varkappa\rho\E^{\I\pi/4}}{k_2\sin\phi_2}
+\rn \tan\phi_1 -\rs \tan \phi_2 \right),
\end{equation}
where $\tan\phi_1 = s\sin\phi_2\left/\sqrt{s_2^2-s^2\sin^2\phi_2}\right.$
and $\Imp \tan \phi_1 \le 0$ (selected by the requirement $\Imp k_1^y \ge 0$).
The reflection coefficient is therefore
\begin{equation}
R_{22}
=
\left|
\frac{\rho \varkappa\E^{\I\pi/4}
+\rn k_2\sin\phi_2\tan\phi_1 -\rs k_2\sin\phi_2 \tan \phi_2}{\rho
\varkappa\E^{\I\pi/4} 
+\rn k_2\sin\phi_2 \tan\phi_1 +\rs k_2\sin\phi_2 \tan \phi_2}
\right|^2.
\end{equation}
These functions for sample parameters are plotted on Fig.\ref{R2}.

\section{Discussion}
It is shown that sound reflection at slanted incidence by a plane
impervious wall is suppressed for both first and second sound. This
phenomenon is similar to Konstantinov effect in ordinary gases.

Coincidentally with the reflection suppression a sound conversion takes
place. The effect is predicted to have strong angle dependence and
should allow experimental verification. Moreover, there is a number of
the heat pulse propagation measurements (\eg \cite{pellame} and
\cite{mulders}) where the pulse transit time was often much shorter than
that for the second sound. This phenomenon is usually explained by
anomalously long phonon free path at low temperatures or by sound
conversion in bulk (due to nonlinear effects) or at liquid-vapour
interface. It seems probable that fast propagation is in fact the
manifestation of the sound conversion described in this paper, so that
the heat pulse is transformed at some wall into the pressure pulse and
is later transformed back near the receiver. The signal therefore
travels (some part of) the path with the velocity of the first sound.

\section*{Acknowledgements}
I thank A.F.\,Andreev and V.I.\,Marchenko for fruitful discussions.
The work was
supported in parts by RFBR grants 06-02-17369, 06-02-17281 and RF
president program 7018.2006.2.

\end{document}